# NONZERO RESULT OF MEASUREMENT OF ACCELERATION OF FREE FALLING GYROSCOPE WITH THE HORIZONTAL AXIS

A. L. Dmitriev, E. M. Nikushchenko, S. A. Bulgakova

*St-Petersburg State University of Information Technologies, Mechanics and Optics*
*49, Kronverksky Prospect, St-Petersburg, 197101, Russia*
*Tel/Fax+7.812.3154071, alex@dmitriyev.ru*

**Resume:** In our experiment we measured the free fall accelerations of the closed container inside which a mechanical rotor (gyroscope) with a horizontal axis of rotation was installed. There was observed an appreciable, essentially exceeding errors of measurements increase of acceleration of free falling of the container at angular speed of rotation of a rotor up to 20 000 rpm.

To laboratory weighings of rotors of mechanical gyroscopes the set of works [1, 2] is devoted. Such measurements were usually carried out with the purpose of experimental check of a equivalence principle, or various gravitoelectric (gravitomagnetic) models. In most cases, in these experiments the axis of a rotor was oriented vertically and, as a whole, the positive effect was absent. In our paper [3] the results of exact weighing of two coaxial rotors with a horizontal axis and with the zero total moment $J_\Sigma$ are given, and its weights which have shown little change, dependent on angular speed of rotation of a rotor. The explanation of these results the possible precession a gyroscope is complicated, connected to rotation the Earth, which essentially could to influence indications of weights, owing to inexact performance of equality $J_\Sigma = 0$. In much smaller degree the precession effects influence on results of measurement of size of acceleration by freely falling of rotor. Thus physical conditions of interaction of a falling rotating rotor with the centre of gravitation (Earth) essentially differ from conditions of weighing of a rotor on based laboratory weights.

In described experiment the acceleration of free falling of container with the two, located coaxially, rotors of mechanical gyroscopes placed inside it was measured; the device and characteristics of the container are given in [3]. On the container the compact highly stable generator of pulses connected to two differ-coloured light-emitting diodes, located along a trajectory of falling of the container is fixed. Distance on centre to centre of aperture stop (holes), established before light-emitting diodes is $l = 76.25 mm$, frequency of impulses $F = 56.25 Hz$, duration of impulse optical signals $0.13 ms$. The trajectory of the falling container was photographed by the digital camera with exposure $0.6 - 0.8 s$ and coordinates of marks (the centres of holes) were digitized by computer.

The calculation of acceleration $g$ of free falling container was carried out under the formula

$$g = \frac{(\Delta_2 - \Delta_1)F^2}{N^2},$$

where $\Delta_1, \Delta_2$ - absolute lengths of the next sites of the trajectories, containing $N$ marks; the scale of the image was defined by distance $l$ between light-emitting diodes. For reduction of influence of aberration of the image owing to distorsion, the average scale of the image paid off on three readout of length $l$ - in the top, central and bottom parts of a trajectory. The size $\bar{g}$ in separate measurement was determined as average value of acceleration, designed on two trajectories appropriate to two groups of color marks on the image.

The example of the measured values of acceleration of free falling container in conditions (1) $\omega = 0$, (2) $\omega \neq 0$ and (3) $\omega = 0$ (upon termination of time rotation of rotor) is shown in figure.

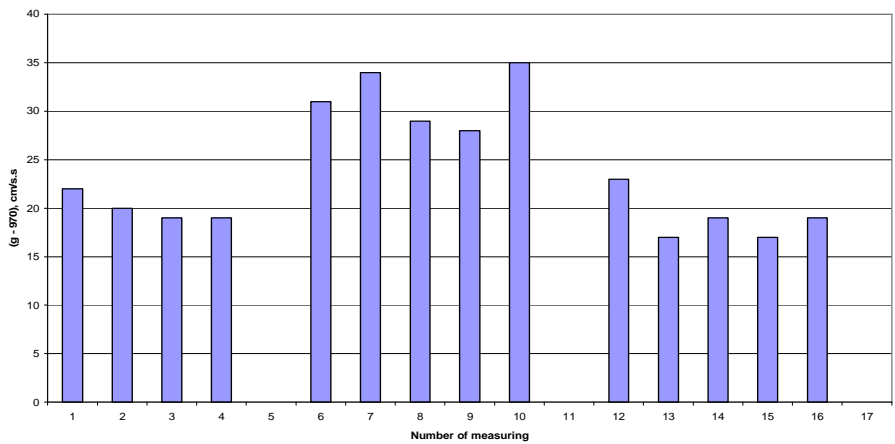

Fig. The example of the measured values of acceleration of free falling container. 1(n. 1-4) $\omega = 0$, 2(n. 6-10) $\omega \neq 0$, 3(n. 12-16) $\omega = 0$.

The maximal angular speed of rotation of a rotor $\omega \approx 20000 rpm$, rotation time of rotor is 14-15 mines, duration of one cycle of measurements from 4-5 pictures about 2 minutes.

It was processed over 200 pictures, thus the increase of acceleration of free falling of a rotor was regularly observed at transition from a condition (1) to a condition (2) with average size $\Delta \bar{g} = 10 \pm 2 cm/s^2$. At smooth reduction of speed $\omega$ of rotation of a rotor the size $\Delta \bar{g}$ also decreased, falling up to zero at $\omega = 0$. In the specified in figure measurements both rotors rotated in one direction and the maximal full moment of rotation of rotors was equaled $J_\Sigma \approx 0.2 kg \cdot m^2/s$. At rotation of rotors in counter directions when $J_\Sigma \approx 0$, small reduction of size $\bar{g}$ was observed.

The reason of an appreciable divergence of the measured absolute value of acceleration $\overline{g}$ of a gravity at $\omega = 0$ (about 990 cm / s$^2$) and standard, at latitude of Saint Petersburg (about 982 cm / s$^2$), apparently, are discrepancies in display of scale $l$, errors of absolute value of frequency $F$ of the generator and also the small local (technical) changes of $\overline{g}$. Geographical orientation of a vector of the moment of rotation of a rotor, N-S or W-O, did not influence on results of measurements $\Delta\overline{g}$. Daily dependence of size $\Delta\overline{g}$ also it was not observed.

At horizontal orientation of an axis of rotation of a rotor the each of its particles simultaneously participates in two linear oscillation in horizontal and vertical planes. Thus of acceleration of particles at their vertical oscillations by an infinite set of derivatives on time from linear displacement are described. As it was marked in [4,5] in these conditions it is possible to expect display of "nonclassical" properties of gravitation which mentioned some more D.Mendeleev [6].

Free falling of masse oscillated along a vertical physically essentially differs from circular (orbital) movement of such masse. Therefore the result received by us does not contradict the results of exact measurements of precession a gyroscope in a circumterraneous orbit.

The further experimental researches of free falling rotating (or oscillated in a vertical plane) masses with application of precision, for example, interferometrical measuring engineering, will promote the deeper understanding of the difficult phenomena of gravitation and inertia.